\documentclass[doublecol]{epl2} 
\newcommand{\e}{\varepsilon}

\title{Convective Lyapunov Spectra}
\shorttitle{Convective Lyapunov Spectra}

\author{Aur\'elien Kenfack Jiotsa \inst{1,2} \and Antonio Politi\inst{2,3,4} \and
 Alessandro Torcini\inst{2,4,5}}
\shortauthor{A. Kenfack \etal}

\institute{                    

\inst{1}
D\'epartement de Physique, Ecole Normale Sup\'erieure, Universit\'e 
de Yaound\'e I, BP 42 Yaound\'e, Cameroon\\
\inst{2} Istituto dei Sistemi Complessi, CNR, via Madonna del Piano 10,
I-50019 Sesto Fiorentino, Italy\\
\inst{3} Institute for Complex Systems and Mathematical Biology, King's
College, University of Aberdeen, Aberdeen AB24 3UE, United Kingdom\\
\inst{4} Centro Interdipartimentale per lo Studio delle Dinamiche Complesse,
via Sansone, 1 - I-50019 Sesto Fiorentino, Italy\\
\inst{5} INFN Sez. Firenze, via Sansone, 1 - I-50019 Sesto Fiorentino, Italy
}


\abstract{We generalize the concept of convective (or velocity-dependent)
Lyapunov exponent $\Lambda(v)$ to an entire spectrum $\Lambda(v,n)$.
Our results are supported by the consistency between the outcome of
the chronotopic approach [{\it S. Lepri et al. J. Stat. Phys., 82 5/6
(1996) 1429}] and a more direct method. There exists a critical
integrated density $n=n_c$, beyond which the convective exponent exhibits a
discontinuous dependence on the velocity, which originates from the
appearance of multiple branches. This phenomenon can be traced back to
a change of concavity of the so-called {\it temporal} Lyapunov spectrum
for $n>n_c$, which is therefore a dynamical invariant.
}

\begin{document}

\maketitle

\section{Introduction}

The linear-stability analysis of an $N$ dimensional chaotic dynamics typically
amounts to computing $N$ Lyapunov exponents (LEs) $\{ \lambda_l \}$ with
$l=1,\dots,N$ (typically ordered from the largest to the most negative one).
In a spatially extended system, the Lyapunov spectrum depends only on the
integrated density $n=l/L$ ($\lambda(n)$) if the system size $L$ is large
enough \cite{grassberger}. Since this tool does not provide any information
on the spatial propagation of perturbations, a new indicator was introduced
in Ref.~\cite{dk87}, the convective (comoving) Lyapunov exponent (CLE)
$\Lambda(v)$, which quantifies the maximal growth rate of an initially localized
perturbation, in a frame that moves with a velocity $v$.
Later, it was recognized that the CLE can be derived from a general theory,
based on the so-called {\it chronotopic} approach \cite{lpt}, which deals with 
the wider class of perturbations with a spatially exponential
profile $\exp(\mu i)$, where $i$ denotes a discrete spatial variable and $\mu$
is a free parameter. As a result, one can, e.g., define the generalized
{\it temporal} Lyapunov $\lambda(n,\mu)$ spectrum. The CLE $\Lambda(v)$ can be
thereof computed as a Legendre transform of the maximal temporal Lyapunov
exponent, i.e. $\lambda(0,\mu)$ \cite{pt94}.

In systems with left-right spatial symmetry, $\Lambda(v)$ is symmetric around
$v=0$, where it attains its maximum value which coincides with the standard
maximum LE. The largest velocity $v=v_c$ such that $\Lambda(v)\ge 0$ is the
maximal propagation velocity of infinitesimal perturbations. In convectively
unstable systems $\Lambda(0)<0$ and the (positive) maximum convective exponent
is attained for some nonzero velocity. In fact, a somehow similar approach was
developed by Huerre and Monkewitz~\cite{hm} to characterize absolute and
convective instabilities in open flows and more recently extended
by Sandstede and Scheel to deal with generic boundary conditions~\cite{bsas}.

In this Letter we go beyond the computation of the growth rate of the local
amplitude of a perturbation, turning our attention to the evolution of volumes
of generic dimension. The chronotopic approach offers a straightforward way
to determine an entire spectrum of {\it convective} LEs, by extending the notion
of Legendre transform from the maximum exponent (i.e. $n=0$) to generic values
of the integrated density $n$. However, it would be desirable to give a more
direct definition as well.
In principle, the most appropriate starting point for a direct definition of
a spectrum of CLE is the approach developed in \cite{cipriani}, where an
ensemble of linearly independent initial conditions (localized within a window
of size $L$) was freely let evolve to thereby determine volume expansion rates
within the very same window. As a result, it was noticed that meaningful
generalized Lyapunov exponents could be defined by simultaneously letting the
time $T$ and the size $L$ tend to infinity, with constant ratio $g=T/L$. The
standard Lyapunov spectrum is recovered for $g\to 0$, while in the
opposite limit $g\to\infty$, all exponents coincide with the maximum. The
dependence on $g$ expresses the fact that the growth along different directions
is affected in a different manner by the local expansion and by diffusion.
In principle, this approach can be implemented for moving windows, too; however,
the unavoidable dependence of the additional parameter $g$, reduces the appeal
of such a direct method.
Accordingly, we have preferred to complement the moving-window approach with
some boundary conditions to get rid of the difficulty of dealing with an
open system. In practice, this is the setup proposed in Ref.~\cite{mkk}.

Altogether, in this Letter we compare this latter method with the chronotopic
approach, finding that they are mutually consistent. Moreover, the chronotopic
approach is by and large the most accurate and this has allowed discovering
serious numerical difficulties that easily lead to artifacts in the direct
computation of the convective Lyapunov spectra (even in the simple case of
coupled maps, herein investigated).
Moreover, the chronotopic approach has revealed that the lower part of the
convective spectrum is charactrized by the existence of multiple solutions. The
phenomenon appears for $n>n_c$ and is associated to
a change of concavity in the temporal Lyapunov spectrum. The critical density
$n_c$ is, by construction, a dynamically invariant dimension density
(analogous to the dimension density of the stable manifold or to the
Kaplan-Yorke dimension density), but the physical meaning of $n_c$ still
needs to be clarified.

Altogether the consistency between the two methods confirms the conjecture that
the chronotopic approach ``encodes" all stability properties of one-dimensional
spatio-temporal systems, that are eventually contained in the corresponding
entropy potential \cite{lpt}.

\section{Theory}

First, we introduce the proper formalism with reference to a standard model of
coupled maps,
\begin{equation}
y^i_{t+1} = f\left [ (1-\e)y^i_{t} + \frac{\e}{2}(y^{i-1}_t+y^{i+1}_t) \right] ,
\label{eq:map}
\end{equation}
where $i=1,\dots,N$ and $t$ are the spatial and temporal (integer) indices, respectively, 
while $\e \in (0,1)$ represents the diffusive coupling and
$f(y)$ is a map of the unit interval onto itself.
We start by investigating the evolution of an infinitesimal
perturbation $\delta y ^i_t$, initially localized in a finite region of length
$M+1 << N $ centered around the origin
($\delta y^i_0 = \xi^i\Theta(M/2-i)\Theta(i+M/2)$, where $\Theta$ is the
Heaviside function and the $\xi^i$'s are iid random variables). 
The standard convective Lyapunov exponent is defined as
\begin{equation}
\Lambda(v,0) = \lim_{t\to\infty} \frac{1}{t} 
\log \frac{|\delta y^{vt}_t|}{|\delta y^{0}_0|}
\label{eq:convm}
\end{equation}
where the ``0" in parentheses signifies that we refer to the maximum
exponent. The formula can be easily extended to generic dynamical systems, by
replacing the absolute value $|\delta y^{i}_t|$ with any norm quantifying the
amplitude of the perturbation on the site $i$ at time $t$. The question
addressed in this Letter is the extension of
this definition to quantify not only the growth rate of the amplitude but also
the expansion/contractions along the additional directions. In order to address
this issue, it is necessary to explore the spatial structure of the
perturbations. In the standard approach, it is sufficient to first let the
perturbation evolve freely in the space and, afterwards, to focus the attention
on specific world lines ($i=vt$). Willing to go beyond the maximal convective
LE, one should consider a set of localized but linearly independent initial
conditions. Accordingly, a theoretical and practical difficulty emerges: that
of defining a proper orthogonalization strategy to avoid the convergence
towards the same direction. In the case of the standard Lyapunov spectrum, this
problem is solved by resorting to the Gram-Schmidt technique \cite{Benettin}. 
However, here it
is not obvious how to do that. The natural idea of referring to the space
covered by the perturbations at a given time leads to the unavoidable
conclusion that one would measure only the tip of the spectrum, since we would
have to rescale by a growing coefficient. Therefore, one is led to restrict
the orthogonalization to a moving window of fixed size $L<M+1$, so that the
computation of convective Lyapunov spectra for different velocities would
require performing separate simulations. This idea, which was already suggested
in \cite{mkk}, is not the
optimal solution, as the initial perturbations are not let to freely evolve,
but artificial modifications have to be introduced on the boundaries to
confine the evolution within the prescribed window. Nevertheless, since it
turns out that boundary conditions do not matter for large enough $L$ 
(see below), we can at least claim that the approach is meaningful.

Let us now be more specific and explain how the LE can be determined in a frame
that moves with a generic velocity $v$ in a discrete spatio-temporal lattice.
We introduce two types of ``moves" in tangent space: the first corresponds to a
static window ($\bf 0$); the second to a right-shift by a single site ($\bf 1$,
without loss of generality we discuss only windows moving to the right).
The corresponding rules are,
\begin{equation}
\delta y_{t+1}^{k} = m_t^i \left [ \frac{\e}{2} \delta y_t^{j-1} +
   (1-\e)\delta y_t^j + \frac{\e}{2}\delta y_t^{j+1}  \right] 
\label{eq:tan0}
\end{equation}
where $1\le k\le L$, $i$ denotes the absolute position in the lattice and
$m_t^i = f'[ (1-\e)y_t^i + \e(y_t^{i-1}+y_t^{i+1})/2]$.
$\bf 0$- and $\bf 1$-iterations correspond to $j=k$ and $j=k+1$, respectively.
The restriction of the rule to a finite interval requires extra assumptions
for $\delta y_t^{L+1}$ and $\delta y_t^0$ ($\delta y_t^{L+2}$), to close the
model for a $\bf 0$ ($\bf 1$) move. We have typically worked by assuming all of
them to be equal to zero, but we have initially verified that the same results
are obtained also for different choices (e.g., $\delta y_t^0 = \delta y_t^1$).
As a last detail, it is necessary to specify the sequence of $\bf 0$s and
$\bf 1$s that is used to study the velocity $v$. Given that $v$ is obviously
equal to the fraction of $\bf 1$s, we have typically selected (simple) rational
numbers, choosing the most uniform sequence with a fixed density
(e.g., for $v=2/5$, $\bf 01010010101001\ldots$).
As a result, for any given velocity $v$, we can determine the whole spectrum
$\Lambda(v,n)$, where $n=l/L$ and $l$ means we refer to the $l$th largest
exponent. 

An alternative approach consists in considering the evolution of a
perturbation with an exponential profile, namely
$\delta y_t^i = \Phi_t^i {\rm e}^{\mu i}$. This requires studying the evolution
equation
\begin{equation}
\Phi_{t+1}^{i} = m_t^i \left [ \frac{\e}{2}{\rm e}^{-\mu}\Phi_t^{i-1} +
   (1-\e)\Phi_t^i + \frac{\e}{2}{\rm e}^{-\mu}\Phi_t^{i+1}  \right] \, ,
\label{eq:tan1}
\end{equation}
where $i=1,\dots,N$ and periodic boundary conditions for $\Phi_t^i$ can be safely assumed.
By solving the model (\ref{eq:tan1}) for different values of $\mu$, one 
obtains the temporal spectrum $\lambda(n,\mu)$. 

We claim that the convective Lyapunov spectrum can be obtained by
a Legendre transform of $\lambda(n,\mu)$,
\begin{equation}
\Lambda(v,n) = \lambda(n,\mu) - \mu v
\label{eq:legen}
\end{equation}
where $v= d\lambda/d\mu$ (and the set of transformations is completed by
the symmetric expression $\mu = d\Lambda/dv$). Altogether Eq.~(\ref{eq:legen})
generalizes to $n>0$ the transformation that has been proved for $n=0$
\cite{pt94}. Its original justification is based on the observation that the
profile of an initially localized perturbation is, at time $t$, locally
exponential with a decay rate $\mu$ that depends on the position $i=vt$.
For $n>0$, it is not clear which profile one should refer to. For this
reason the above definition is essentially a formal one.

\section{Numerical Results}

We have tested both approaches, by studying a chain of coupled
logistic maps, i.e. with $f(y) = ry(1-y)$ ($r=4$), $y \in [0;1]$ and $\e=1/3$
(these are  the same parameter values adopted in Ref.~\cite{mkk}).

\begin{figure}
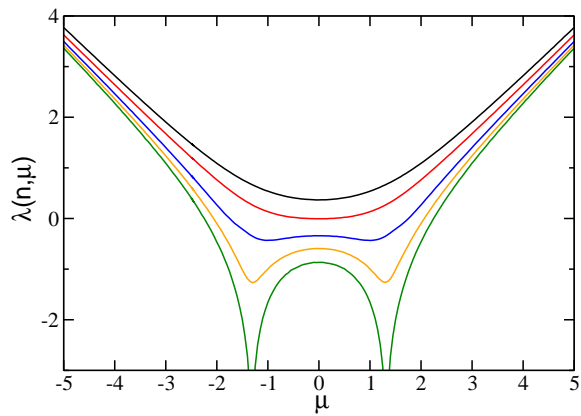

\onefigure[draft=false,clip=true,height=0.3\textwidth]{fig1.eps}
\caption{(Color Online) Temporal Lyapunov exponents $\lambda(n,\mu)$ versus 
$\mu$ for a chain of $N=100$ coupled logistic maps with $r=4$ and $\e=1/3$. From
top to bottom the curves refer to $n =0$, 0.5, 0.75, 0.9 and 1.
}
\label{fig:1}
\end{figure}

In Fig.~\ref{fig:1} we plot the temporal spectra for $N=100$ and some values of
$n$ (namely, $n=0$, 0.5, 0.75, 0.9 and 1). We have also verified that $N$ is
large enough to ensure the thermodynamic limit. The uppermost curve corresponds
to the maximum exponent that is known to grow monotonously for increasing (in
absolute value) $\mu$. On the other hand, the lowermost curve, which corresponds
to the minimum LE not only is non monotonous, but even exhibits a singular
behavior for $|\mu | \approx 1.31$. 

\subsection{Convective exponents vs velocity}

\begin{figure}
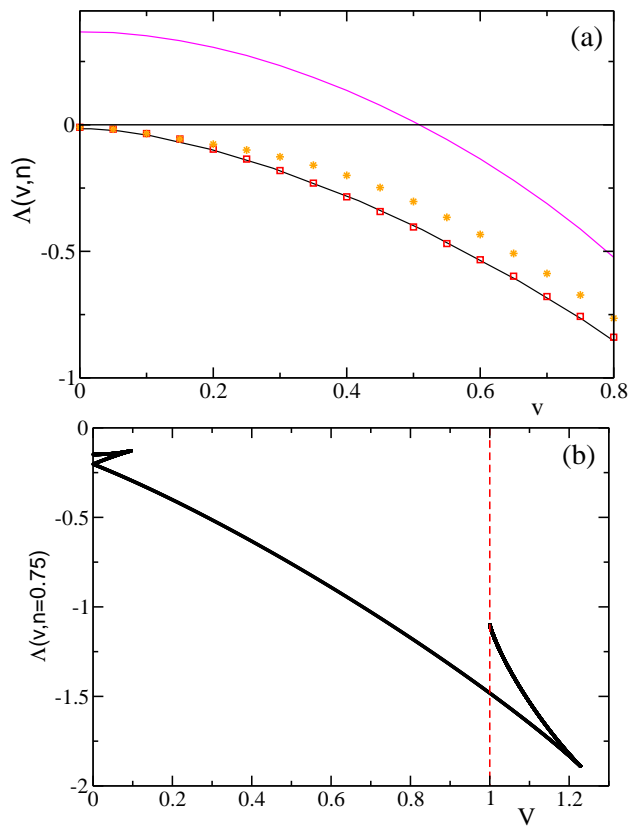

\onefigure[draft=false,clip=true,height=0.3\textwidth]{fig2a.eps}
\onefigure[draft=false,clip=true,height=0.3\textwidth]{fig2b.eps}
\caption{(Color Online) Convective exponents $\Lambda(v,n)$ vs. the velocity of
the comoving reference frame for Logistic (a) and Bernoulli (b) coupled
map lattices. (a) The convective exponents have been obtained for
$n=0$ (upper curve) and $n=0.5$
(lower curve). The solid lines have been obtained be Legendre transforming
(see Eq.~\ref{eq:legen}) the curves reported in the previous figure.
The symbols correspond to direct estimates of the convective exponents in
double precision for $n=0.5$ and $L=100$:
(orange)  asterisks  refer to $\gamma=0$, while (red) squares to $\gamma=1.0$.
(b) The reported convective spectrum has been analytically determined 
by performing the Legendre transform of Eq. (\ref{lambda_bern}) for $n=0.75$
and $N=100$.}
\label{fig:2}
\end{figure}

In Fig.~\ref{fig:2}a we plot the convective spectra obtained
by Legendre transforming two of the curves reported in  Fig.~\ref{fig:1}.
The upper solid curve is the standard convective spectrum. It starts from
a maximum value for $v=0$ that corresponds to the usual maximum LE and
crosses the zero axis at $v_c=0.510(5)$ which indicates the maximal propagation
velocity of perturbations. The lower solid curve corresponds to $n=0.5$
and the very fact it is nearly zero for $v=0$ indicates that approximately
half of the standard LEs are positive.

In order to check the equivalence between the direct and the chronotopic
approach, we have computed the convective spectra also by iterating localized
perturbations, implementing the method described in the previous section.
The (orange) asterisks in Fig.~\ref{fig:2}a correspond to the outcome of 
simulations performed with $L=100$. The agreement is rather poor and does even get worse upon
increasing the window size $L$ (not reported). However, we discovered that the
problem is not a conceptual, but a numerical one. The reason is that the
Lyapunov vectors in the moving windows are exponentially localized around the
left border of the window. Accordingly, many of their components are so small
that double-precision accuracy (i.e. with 14-15 digits) is not sufficient,
especially for large values of $L$. As a matter of fact we have verified that
by employing extended precision (i.e. with approximately 30 digits) the
agreement increases. However, an even better agreement can be obtained
already with double precision, by using weighted Euclidean.
More precisely, given any two vectors
${\bf u}= \{ u^i \}$, ${\bf v}= \{v^i\}$, we define the scalar product as
\begin{equation}
{\bf u} \cdot {\bf v} = \sum_i u_i v_i {\rm e}^{\gamma i}
\end{equation}
where $\gamma$ is a free parameter. Formally speaking, it is well known that 
Lyaphnov exponents are independent of the norm that is chosen, i.e., in this
case, of $\gamma$. Nonetheless, the numerical accuracy may significantly depend
on $\gamma$. In fact, by comparing double with extended precision for different
values of $\gamma$ we concluded that $\gamma=1$ is a nearly optimal choice
\footnote{At least for not too large velocities, otherwise the choice of the
norm does not help enough}.
The data reported in Fig.~\ref{fig:2}a (red squares) indeed confirm the 
increased agreement with the chronotopic results.
However, a proper selection of $\gamma$ does not solve completely the problem:
for large velocities and larger values of $L$, numerical accuracy remains a
serious issue.

Anyway, the relevant message that comes from the numerical analysis is that
the definition of convective Lyapunov spectra through the chronotopic
approach is not just formal but provides the correct answer and, moreover,
the method is far more reliable than the direct one, since one does not
face accuracy problems. Incidentally, it is the comparison between the two
approaches that has allowed discovering the serious problems of numerical
accuracy that affect the direct approach and thereby the simulations reported
in Ref.~\cite{mkk}.
Last but not least, for large values of the integrated density $n$,
the non-monotonicity of $\lambda(n,\mu)$ leads to the emergence of
extra branches in the convective spectrum, as shown in Fig.~\ref{fig:2}b
(which refers to Bernoulli maps $f(x) = 2x$ ${\rm mod}(1)$ with the same coupling
constant as before).
The origin of the three branches at small velocities will be addressed in the
following section, while the coexistence of branches at large velocities
is totally irrelevant. In fact, it arises in an unphysical region,
since $v>1$ cannot be obtained in a lattice with nearest neighbor coupling.
We interpret the very existence of the two branches as an instance of a
phase velocity that can be faster than ``light velocity" without causing
any paradox.

\subsection{Convective exponents vs density} 
\begin{figure}
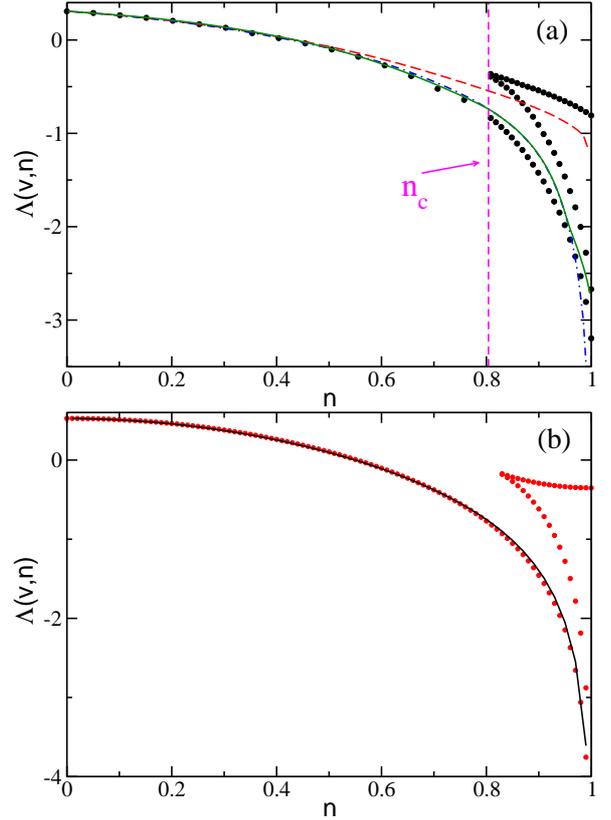

\onefigure[draft=false,clip=true,height=0.3\textwidth]{fig3a.eps}
\onefigure[draft=false,clip=true,height=0.3\textwidth]{fig3b.eps}
\caption{(Color Online) Spectrum of the convective exponents $\Lambda(v,n)$. (a)
Coupled logistic maps for $v=1/5$. The solid lines correspond to direct
measurements. From top to bottom: $L=100$ and $\gamma=0$
(red dashed line); $L=200$ and $\gamma=2$ (blue dot-dashed line);
$L=100$ and $\gamma=2$ (green solid line). Full circles correspond to the Legendre transform for
$N=100$. Finally, the vertical dot-dashed (magenta) line indicates the
$n_c$-value. (b) Bernoulli maps for $v=1/3$, estimated analytically via the
Legendre transform (red circles) and directly by diagonalizing a constant
$50 \times 50$ matrix (solid line) with extended precision.
}
\label{fig:3}
\end{figure}

We now compare the convective
spectra obtained for a given velocity, as it helps
clarifying the behavior for large integrated densities. An example is reported
in Fig.~\ref{fig:3}a for $v=1/5$. Above a certain density $n_c$ the spectrum
obtained via the Legendre transform displays three different branches:
the lower one is associated to positive $\mu$-values, while the other
two to negative $\mu$. Once again the standard direct estimates suffer problems
of numerical accuracy (see the upper solid curve in Fig.~\ref{fig:3}a). The
improved simulations (for $\gamma=2$) seem to converge towards the lower branch,
but there is still some discrepancy and it is difficult to decide whether
this is due to finite-size corrections.

In order to obtain a more convincing evidence, we now consider a model of
Bernoulli maps. Since the multipliers are constant in space
and time, we expect smaller finite size corrections and, moreover, the
estimation of the convective spectra is simpler as it reduces to the
diagonalization of a matrix (for rational velocities). In fact, in this model,
the tangent matrices (see Eq.~(\ref{eq:tan0})) depend only whether the move is of
type $\bf 0$ or $\bf 1$.
Therefore, given a rational velocity, characterized by a periodic sequence of
$\bf 0$s or $\bf 1$s (with period $P$), it is sufficient to multiply $P$ of such
matrices and thereby determine the eigenvalues. The results for
$v=1/3 =(001001001\ldots)$ are reported in Fig. \ref{fig:3}b, where one can
appreciate that the agreement between the lower branch and the direct
method  is already quite impressive for a window of size $L=50$. 

\begin{figure}
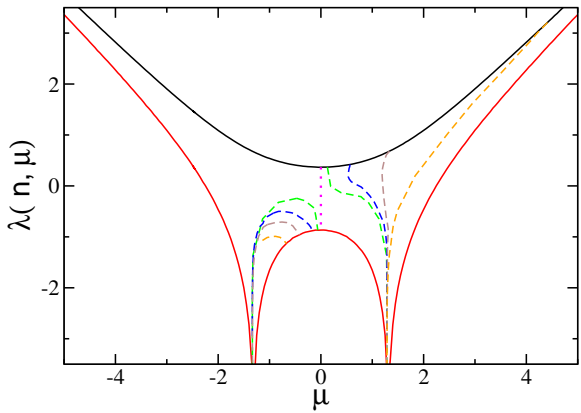

\onefigure[draft=false,clip=true,height=0.3\textwidth]{fig4.eps}
\caption{(Color Online) Isolines in the plane $(\lambda,\mu)$ corresponding
to a constant velocity $v$ for coupled logistic maps. The vertical dotted
line refer to $v=0$, the other dashed lines from left to right for $\mu >0$ 
(resp. right to left for $\mu <0$) correspond to $v=0.05$, 0.20, 0.50 and
0.95 with $N=100$.}
\label{fig:5}
\end{figure}

In order to shed some light on the origin of these branches, we have plotted 
the paths in $(\lambda,\mu)$-plane that correspond to spectra with different
velocities (see Fig.~\ref{fig:5}). They are the isolines where the
slope $d\lambda/d \mu=v$ stays constant. The vertical line at $\mu = 0$ is the
path for the standard Lyapunov spectrum. However, for finite velocities
the line breaks into two components that lie in the positive and negative
$\mu$ half-planes, respectively: the former one corresponds
to the lower branch in Fig.~\ref{fig:3}, while the latter one gives rise to the
two upper branches. 
Altogether it is reasonable to conjecture that the reason for
discarding the upper branches is that they correspond to negative $\mu$ values,
i.e. to profiles that are larger on the right side. In fact, this is impossible
unless one assumes the presence of some source of ``noise" on the right
of the moving window. 

A moment's reflection suggests that the mathematical origin of two branches
in Fig.~\ref{fig:5} is the change of concavity of $\lambda(n,\mu)$
(as a function of $\mu$) upon increasing $n$: it suddenly enforces points
with positive derivative to jump from the right to the left side of $\mu=0$.
Once again the Bernoulli maps allow for an accurate investigation.

\begin{figure}
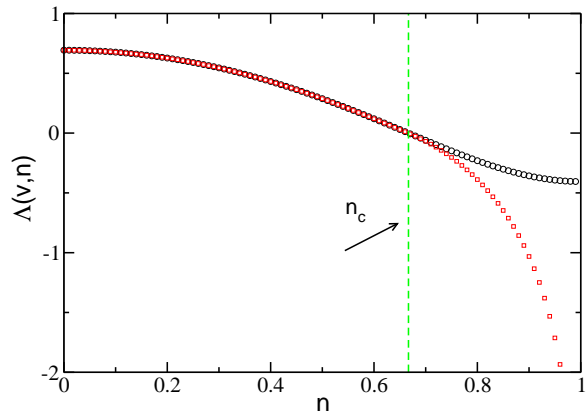

\onefigure[draft=false,clip=true,height=0.3\textwidth]{fig5.eps}
\caption{(Color Online) Spectrum of the convective exponents $\Lambda(v,n)$ for
coupled Bernoulli maps for $v=0$ (circles) and $v=0.001$ (squares), estimated
by Legendre transforming of Eq.~(\ref{lambda_bern}).
The parameters are $a=2$ and $\varepsilon=1/3$.}
\label{fig:6}
\end{figure}

The expression for the temporal exponent is \cite{lpt},
\begin{eqnarray}
&& \lambda(n,\mu)=\log a+\frac{1}{2}\log\left|(1-\varepsilon)^2+ \right.
\label{lambda_bern}
\\
&& \left. 2\varepsilon(1-\varepsilon)
\cosh \mu\cos \pi n+\varepsilon^2(\cosh^2 \mu -\sin^2 \pi n)\right|
\nonumber
\end{eqnarray}
By expanding this expression for small values of $\mu$, it is easy
to verify that the quadratic term in $\mu$ changes sign for
\begin{equation}
n = n_c \equiv 1 - \frac{{\rm acos} [\e/(1-\e)]}{\pi} \, ,
\end{equation}
provided that $\e<1/2$.
As a matter of fact we have estimated the spectrum for the convective
exponents for $v=0$ and $v=0.001$. In  Fig.~\ref{fig:6}. one can appreciate
that for $n<n_c$ they are nearly identical, while for $n>n_c$, they separate
out by a finite amount as a result of the selection of the lower branch.
Now, we can return to $\Lambda(v,n)$ as a function of $v$ for fixed
$n$ (illustrated in Fig.~\ref{fig:2} for $n=0.5$) and conclude that 
$\Lambda(v,n)$ must have a discontinuity in $v=0$, if $n>n_c$.

\section{Discussion and conclusions}
In this Letter, we have shown that the notion of convective Lyapunov exponent
can be extended from the maximum to an entire spectrum. The comparison between
two different methods reveals a rather complex scenario with several issues
that need be further clarified. An example is the dimension density $n_c$ which
separates a standard behavior from the appearance of multiple branches.
Its very existence is connected to a change of concavity in the temporal
Lyapunov spectra. We have verified that in coupled maps it generally exists for
not-too-large coupling ($\e<1/2$ in Bernoulli maps) and preliminary simulations
confirm the existence of such a critical density also in Stuart-Landau
oscillators \cite{KPT}.
Nevertheless, its physical meaning is rather unclear. We can only claim that
the dimension $n_c$ is a dynamical invariant (as it follows from a general
property of the Lyapunov spectrum) like the density of unstable directions
(fraction of positive Lyapunov exponents), the Kaplan-Yorke dimension and
the dimension of physical modes \cite{chate}.
Another question concerns the upper branches that we have dismissed as
irrelevant, but could play some role in specific physical contexts.

Finally, the direct definition is still partially unsatisfactory, as
it involves the addition of in-principle-unnecessary boundary conditions. An
open system approach such as that developed in \cite{cipriani} would be much
more appealing, but it requires incorporating the additional parameter
$g=T/L$ in the current theory.

An alternative idea could be that of referring to covariant Lyapunov
vectors \cite{ginelli}: if one could indeed ``build" the initial perturbation
by using only such vectors, one would be automatically assured that
no more-unstable degrees of freedom are going to be excited.
The problem is that one first needs to evolve back and forwards the chain
on sufficiently long time scales to allow the perturbation propagate over
sufficiently large distances to measure asymptotic quantities.

\acknowledgments
A.K.J. undertook this work with the support of 
the ICTP ``Programme for Training and Research 
in Italian Laboratories'' (TRIL), Trieste, Italy.
AT acknowledges the Villum Foundation for partial support
received during his visit to the Department of Physics and Astronomy
of the University of Aarhus (Denmark).

\end{document}